\documentclass[conference]{IEEEtran}
\usepackage{amssymb}
\usepackage{amsmath}
\usepackage{amsfonts}
\usepackage{graphicx}
\usepackage{algorithm}
\usepackage{algorithmic}
\usepackage{cite}
\usepackage{epstopdf}
\usepackage{color}
\usepackage{multirow}
\usepackage{subfigure}
\usepackage{bm}
\usepackage{lipsum}
\usepackage{multicol}

\IEEEoverridecommandlockouts
\begin{document}
\title{Iterative detection and decoding for SCMA systems with LDPC codes}
\author{Baicen Xiao, Kexin Xiao, Shutian Zhang, Zhiyong Chen, Bin Xia and Hui Liu \\
Department of Electronic Engineering, Shanghai Jiao Tong University, Shanghai, P. R. China\\
Email: {\{xinzhiniepan, kexin.xiao, zhangshutian, zhiyongchen, bxia, huiliu\}@sjtu.edu.cn}
}
\maketitle

\begin{abstract}
Sparse code multiple access (SCMA) is a promising multiplexing approach to achieve high system capacity. In this paper, we develop a novel iterative detection and decoding scheme for SCMA systems combined with Low-density Parity-check (LDPC) decoding. In particular, we decompose the output of the message passing algorithm (MPA) based SCMA multiuser detection into intrinsic part and prior part. Then we design a joint detection and decoding scheme which iteratively exchanges the intrinsic information between the detector and the decoder, yielding a satisfied performance gain. Moreover, the proposed scheme has almost the same complexity compared to the traditional receiver for LDPC-coded SCMA systems. As numerical results demonstrate, the proposed scheme has a substantial gain over the traditional SCMA receiver on AWGN channels and Rayleigh fading channels.
\end{abstract}
\section{Introduction}
Sparse code multiple access~(SCMA) has attracted much attention since it is capable of  supporting massive connections simultaneously, yielding a competitive candidate for Fifth Generation (5G) communications\cite{nikopour2013sparse}. Commonly, SCMA can be viewed as a generalization of sparse spread CDMA\cite{guo2008multiuser}, with a few numbers of nonzero elements within a signature. For an uplink SCMA system, when each user is assigned a specific codebook, the multiplexing becomes a superposition scheme which will obtain the shaping gain. However, on the receiver side serious multiple address interference (MAI) is the main obstacle to implement multiuser detection. The optimum maximum a posterior (MAP) algorithm obviously shows the best performance with considerable complexity. In order to tackle the high complexity of MAP algorithm, some low complexity algorithm are proposed to handle this NP-complete problem\cite{verdu1998multiuser} within tolerable performance loss. Especially, thanks to the sparse structure of SCMA, the complex MAP formula can be solved iteratively with sum-product algorithm or message passing algorithm (MPA)\cite{kschischang2001factor}.

 Lately, in order to improve the bit error rate (BER) performance of SCMA, \cite{huawei} has introduced the Turbo-principle, which is widely used in detection and decoding problems such as joint source-channel coding \cite{hagenauer1997turbo,hagenauer2003turbo} and multiuser detection \cite{wang1999iterative,sanderovich2005ldpc}, to exchange information between the SCMA detector and the channel decoder. However,  the proposed Turbo-like scheme  in \cite{huawei} doesn't take full advantage of the iterative structure of SCMA detection and suffers from high complexity proportional to the number of outer iterations.
For the sake of taking full advantage of the iterative characteristic of SCMA detection, we need a kind of channel coding which applies iterative decoding. Low-Density Parity-Check (LDPC) codes which are excellent error correcting codes providing a large coding gain \cite{richardson2003renaissance} and adopt iterative decoding, are especially suitable for our requirments.

The goal of this paper is to apply a novel Turbo-like combination of SCMA multiuser detection and LDPC decoding. The difference from \cite{huawei} should be noted that this paper do a novel Turbo-like combination of iterative detection and iterative decoding, i.e, during each outer loop only partial inner iterations in detector and decoder are implemented, to obtain a satisfied performance gain with almost the same complexity compared to the receiver without Turbo-like scheme. Firstly, we investigate in detail the MPA-based SCMA multiuser detector from the perspective of solving marginal function and then deduce the SCMA multiuser detection algorithm in logarithmic form. Furthermore, the intrinsic information is decomposed from the output of both SCMA multiuser detector and LDPC decoder and the way intrinsic information interacts between detector and decoder is presented. As numerical results show, this scheme achieves a 0.9 dB performance gain in terms of BER with almost the same complexity for both AWGN channels and Rayleigh fading channels compared to traditional receiver for LDPC-coded SCMA systems.

For the sake of clarity, throughout this paper, the sets of binary and complex numbers are denoted by $\mathbb{B}$ and $\mathbb{C}$, respectively. Upper-case calligraphic symbols $\mathcal{X}$ denote constellation sets and log$(\cdot)$ denotes natural logarithm. To represent a scalar, a vector and a matrix, we use $x$, $\mathbf{x}$ and $\mathbf{X}$, respectively.

\section{System model}
\begin{figure*}[t]
            \centering
            \includegraphics{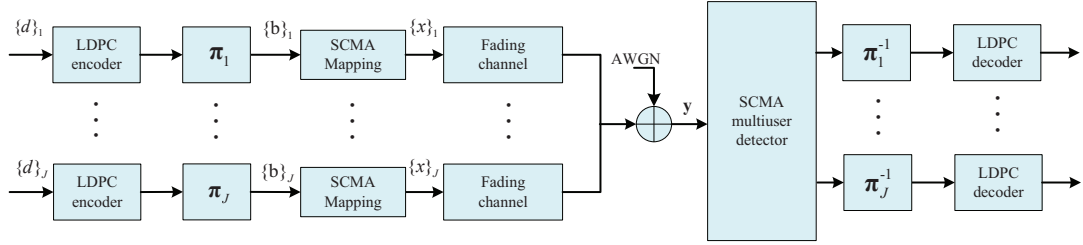}
            \hspace{2in}\parbox{1\linewidth}
{\caption{Block diagram for uplink SCMA system }
\label{system_diagram}}
            \end{figure*}

\begin{figure}[t]
            \centering
            \includegraphics[width=3.2in]{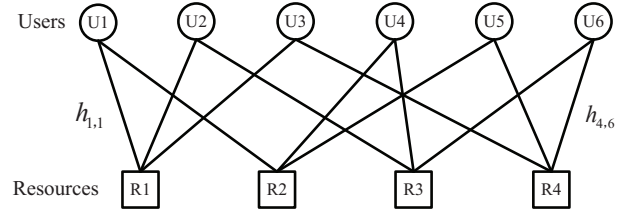}
            \hspace{2in}\parbox{1\linewidth}
{\caption{Factor graph representation for $6$ users and $4$ resources SCMA, $d_j=2$ for all $j$ and $d_k = 3$ for all $k$. The edges between user nodes and resource nodes can be seen as the channels. }
\label{factor-graph}}
            \end{figure}

We consider an uplink LDPC-coded SCMA system with $J$ users and $K$ resources, and signaling through fading channels with additive white Gaussian noise (AWGN), as shown in Fig. {\ref{system_diagram}}.
 For each user $j$, $j$ = 1, $\cdots$ , $J$, data bits \{$d_{n_0}^{j}\mid n_0=1,2,\cdots, n$\} are first encoded into \{$b_{m_0}^{j}\mid m_0=1,2,\cdots, m$\} by an LDPC encoder with code rate $R_j=n/m$. In order to reduce error bursts and take advantage of diversity gain, the coded bits are permuted by an interleaver $\pi_j$. Every log$_2$$(M)$ interleaved coded bits $\{b_{m^{\pi}}^{~j} | m^{\pi}=1,2,\cdots,\text{log}_2(M)\}$ is grouped together and then mapped by SCMA mapper $f_j$ into a $K$-dimensional complex symbol as $f_j$: $\mathbb{B}^{\text{log}_2(M)}\rightarrow{\mathbf{x}}_{j}\in\mathcal{X}^j\subset\mathbb{C}^K$ with cardinality $|\mathcal{X}^j|$ = $M$. Because of the sparsity of SCMA, a $K$-dimensional symbol ${\mathbf{x}}_{j}$ consists of $N_j$ $<$ $K$ non-zero elements, each corresponding to an OFDMA tone or other resources. For the receiver end, the received siganls are the superposition of $J$ users' signals and ambient noise, which can be written in a discrete-form, if memoryless channel considered, as
\begin{equation}
\ \mathbf{y} = \sum_{j=1}^J{\text{diag}(\mathbf{h}_{j}){\mathbf{x}}_j} + \mathbf{n}\text{,}
\end{equation}
where $\mathbf{y} = (y_{1},\cdots,y_{K})^\text{T}$ is the received signal vector, $\mathbf{h}_{j} = (h_{1j},\cdots,h_{Kj})^\text{T}$ is the channel vector for user $j$, $\mathbf{x}_{j} = (x_{1j},\cdots,x_{Kj})^\text{T}$ is the symbol transmitted by user $j$, and $\mathbf{n}$ is a white Gaussian noise vector subject to $\mathcal{CN}(0,N_0\mathbf{I})$. Apparently, from (1) the received signal at resource $k$ can be written as
\begin{equation}
\ y_k = \sum_{j=1}^J{h_{k,j}x_{k,j}}+n_k,  k=1,\cdots, K.
\end{equation}
It is easy to recognize from (2) that each user sees interference from other $K-1$ users. However, since the signal vector from arbitrary user is sparse, i.e., not all users contribute to the k-th resource, the interference is hence reduced and (2) can be re-written as
\begin{equation}
\ y_k = \sum_{j\in{\partial k}}{h_{k,j}x_{k,j}}+n_k,  k=1,\cdots, K.
\end{equation}
where $\partial k$ denotes the users contributing to the $k$-th resource, called the neighborhood of node $k$, and this relationship which is decided by the SCMA mapper can be presented by factor graph and indicator matrix $\mathbf{C}$. Let $d_j$ and $d_k$ be the number of resources occupied by user $j$ and the number of users resource $k$ is connected, respectively. For the sake of clarity, we give an example of SCMA factor graph. Assuming $d_j = 2$ for all $j$ and $d_k = 3$ for all $k$, the factor representation is shown as Fig. {\ref{factor-graph}}, and the corresponding indicator matrix is

\[\mathbf{C}=
\left[
\begin{array}{cccccc}
1 & 1 & 1 & 0 & 0 & 0\\
1 & 0 & 0 & 1 & 1 & 0\\
0 & 1 & 0 & 1 & 0 & 1\\
0 & 0 & 1 & 0 & 1 & 1
\end{array}
\right]
\]
therein, $c_{k,i}=1$ means resource $k$ is occupied by user $i$.

It's observed that this factor graph is regular, i.e., all the user nodes have the same degree, so do the resource nodes. It should be emphasized that this regular structure may not be the best, in another word, it's possible for a irregular structure to play a better performance if there are diversities between the quality of different resources. In this paper, however, our attention is only focused on regular SCMA structure, but some of the results can be easily extended to the irregular scenario and the irregular structure will be our future work.

\section{ Iterative  detection and decoding  }
In this section, we analyse SCMA detection based on MPA iterative algorithm, and propose an effective combination method of SCMA multiuser detection and LDPC decoding.
\subsection{SCMA detection}
Firstly, we describe why the MPA algorithm can be applied to SCMA detection from the perspective of solving marginal function. Given received signal $\mathbf{y}$ and assuming ideal channel estimation, a SCMA detection based on MAP is to choose a matrix $\hat{\mathbf{X}}=(\mathbf{x}_1, \cdots, \mathbf{x}_J)$ to maximizing the joint $a~posterior$ pmf, which is expressed as
\begin{equation}
\  \hat{\mathbf{X}} = \arg\max_{\mathbf{X}\in{\mathcal{X}^{KJ}}} p(\mathbf{X}|\mathbf{y})\text{,}
\end{equation}
where $\mathcal{X}^{KJ}$ denotes the set of all possible symbols, i.e., the $j$-th column of $\mathcal{X}^{KJ}$ is the set $\mathcal{X}^j$ described in Section \uppercase\expandafter{\romannumeral2}. In order to estimate the information of user $j$, we can choose a $\mathbf{\hat{x}}_j$ to maximize the marginal $a~posterior$ pmf with respect to $\mathbf{x}_j$  as
\begin{equation}
\  \hat{\mathbf{x}}_j = \arg\max_{z\in\mathcal{X}^j} \sum_{\substack{\mathbf{X}\in{\mathcal{X}^{K\!J}}\\{\mathbf{x}_j=z}}}{p(\mathbf{X}|\mathbf{y})}\text{.}
\end{equation}
Applying Bayes's rule, we can get
\begin{equation}\label{SCMA_f(x|y)}
\  p(\mathbf{X}|\mathbf{y})=\frac{p(\mathbf{y}|\mathbf{X})p(\mathbf{X})}{p(\mathbf{y})}
\end{equation}
\begin{equation}
\  \propto p(\mathbf{y}|\mathbf{X})p(\mathbf{X})=p(\mathbf{X})\prod_{k=1}^K{p(y_k|\mathbf{X})}\text{.}
\end{equation}
The last equation follows the fact that the elements of noise vector is identically independent distributed (i.i.d.) and uncorrelated with transmitted symbols, hence once the transmitted symbols are given, different dimension of received signal $\mathbf{y}$ are independent.

Since $y_k$ is influenced by parts of users, i.e., users from $\partial k$, equation (5) can be reduced to
\begin{equation} \label{SCMA_x_j}
\  \hat{\mathbf{x}}_j = \arg\max_{z\in\mathcal{X}^j} \sum_{\substack{\mathbf{X}\in{\mathcal{X}^{K\!J}}\\{\mathbf{x}_j=z}}}{p(\mathbf{X})\prod_{k=1}^K{p(y_k|\mathbf{x}_p, p\in{\partial k})}}~~\forall j\text{.}
\end{equation}
Furthermore transmitted symbols from different users are independent, $p(\mathbf{X})$ can be written in a product form as $p(\mathbf{X})=\prod_{q=1}^{J}{p(\mathbf{x}_q)}$, and it's clear that the sum terms in equation (\ref{SCMA_x_j}) can be given by
\begin{equation} \label{SCMA_f(x)}
\  f(\mathbf{x}_1,\cdots,\mathbf{x}_J) = \prod_{q=1}^{J}{p(\mathbf{x}_q)}\prod_{k=1}^K{p(y_k|\mathbf{x}_p, p\in{\partial k})}\text{.}
\end{equation}

Traditionally, to solve equation (\ref{SCMA_x_j}) needs $J$ operations where redundant computing exits. Because of the product form (\ref{SCMA_f(x)}), thanks to the method in \cite{kschischang2001factor}, we can solve a marginal function problem of multivariable function like (\ref{SCMA_x_j}) iteratively based on factor graph, which is usually called sum-product algorithm or MPA algorithm. Each node in the factor graph sends ``belief message" to its neighbors during each iteration and the ``belief message" shouldn't be sent back during the next iteration, hence the inference can be made sufficiently in the graph after some iterations. It should be emphasized that when the factor graph of equation (\ref{SCMA_f(x)}) is cycle-free MPA algorithm is able to produce the accurate marginal function  \cite{guo2008multiuser}, but for the factor graph with cycles this algorithm is suboptimal.

The iterative SCMA multiuser detection algorithm based on MPA is presented in algorithm 1.
\begin{algorithm}
\caption{Iterative SCMA multiuser detection algorithm}
\begin{algorithmic}[1]
\STATE \textbf{Variable definition}\\
       $V_{j\rightarrow k}^{t}(\mathbf{x}_j)$: the message sent from $j$-th user node to $k$-th resource node during $t$-th iteration\\
       $U_{k\rightarrow j}^{t}(\mathbf{x}_j)$: the message sent from $k$-th resource node to $j$-th user node during $t$-th iteration\\
\STATE \textbf{Initialization} \\
        $p(\mathbf{x}_j) \leftarrow \frac{1}{M}$, for all $\mathbf{x}_j\in\mathcal{X}^j$ and $j=1, \cdots, J$.\\
        \textbf{for all} $j=1, \cdots, J$ and $k=1, \cdots, K$ \textbf{do}\\
$U_{k\rightarrow j}^{0}(\mathbf{x}_j) \leftarrow 1 $, for all $\mathbf{x}_j\in\mathcal{X}^j$\\
\FOR{$t=1,t\leq T,t++$}
\STATE \textbf{for~all} $j,k$ with $c_{jk}=1$ and $\mathbf{x}_j \in \mathcal{X}^j$ $\textbf{do}$\\
\begin{equation}
V_{j\rightarrow k}^{t}(\mathbf{x}_j) \leftarrow p(\mathbf{x}_j)\prod_{s\in{\partial j}\backslash k}{U_{s\rightarrow j}^{t-1}(\mathbf{x}_j)}
\end{equation}
\STATE \textbf{for~all} $j,k$ with $c_{jk}=1$ and $\mathbf{x}_j \in \mathcal{X}^j$ $\textbf{do}$\\
\begin{align}
U_{k\rightarrow j}^{t}(\mathbf{x}_j)&\leftarrow \sum_{(\mathbf{x}_p)\partial k \backslash j}\frac{1}{\pi N_0}\text{exp}[\frac{1}{N_0}\parallel y_k-h_{k,j}x_{k,j}\nonumber\\
&-\sum_{p\in\partial k \backslash j}{h_{k,p}x_{k,p}}\parallel ^2]\prod_{p\in{\partial k}\backslash j}V_{p\rightarrow k}^{t}(\mathbf{x}_p)
\end{align}
\ENDFOR
\STATE  \textbf{return} $\textbf{for~all}$ $\mathbf{x}_j\in\mathcal{X}^j$ and $j=1,\cdots,J$\\
$V_j(\mathbf{x}_j)\leftarrow p(\mathbf{x}_j)\prod_{s\in{\partial j}}{U_{s\rightarrow j}^{t-1}(\mathbf{x}_j)}$
\end{algorithmic}
\end{algorithm}
Therein£¬ ``$\partial k \backslash j$'' denotes the neighborhood of node $k$ excluding node $j$ and the notation ``$\sum_{(\mathbf{x}_p)\partial k \backslash j}$" denotes sum over all possible values of $\mathbf{x}_p \in \mathcal{X}^p$ for all $p \in \partial k \backslash j$. In this algorithm, we assume all possible symbols are equiprobable, that is to say there is no prior information. When prior information is provided by channel decoder as described later, $p(\mathbf{x}_j)$ should be updated. For the convenience of implementation, we then derive the SCMA detection algorithm in logarithmic form, given prior information in the form of the log-likelihood ratio. Note that one of our goal is to decompose the output of SCMA detector into intrinsic information and prior information in order to do a novel Turbo-like combination with LDPC decoder in subsection $C$.

Once the prior information $\{L^{s,p}(\mathcal{X}^j)\}_{j=1,\cdots,J}$ is given, here the superscript ``s" denotes the LLR is with respect to symbol, the posterior probability of $\mathbf{X}$ can be written as
\begin{equation}\label{SCMA_f(x|y,l)}
\begin{split}
\  p(\mathbf{X}|\mathbf{y},&\{L^{s,p}(\mathcal{X}^j)\}_{j=1,\cdots,J})=\\
&C\cdot p(\mathbf{y}|\mathbf{X})p(\mathbf{X}|\{L^{s,p}(\mathcal{X}^j)\}_{j=1,\cdots,J})\text{,}
\end{split}
\end{equation}
where $C=\frac{p(\{L^{s,p}(\mathcal{X}^j)\}_{j=1,\cdots,J})}{p(\mathbf{y},\{L^{s,p}(\mathcal{X}^j)\}_{j=1,\cdots,J})}$ is a constant during each inner loop of SCMA detection. Comparing equation (\ref{SCMA_f(x|y)}) with (\ref{SCMA_f(x|y,l)}), it's easy to show that only the initialization part of algorithm 1 should be modified when prior information is available. Then ``$p(\mathbf{x}_j) \leftarrow \frac{1}{M}$" should be replaced by ``$p(\mathbf{x}_j) \leftarrow p(\mathbf{x}_j|L^{s,p}(\mathbf{x}_j))$".
Let us fix a reference point $\mathbf{\tilde{x}}_j$ , which denotes all ``1" transmitted to facilitate the following deduction, for each user $j$. Let $LV_{j\rightarrow k}^{t}(\mathbf{x}_j)$ and $LU_{k\rightarrow j}^{t}(\mathbf{x}_j)$ be the information in logarithmic form from user node $j$ to resource node $k$ and from resource node $k$ to user node $j$, respectively, then,
\begin{equation}\label{LV}
\begin{split}
LV_{j\rightarrow k}^{t}(\mathbf{x}_j) &= \text{log}\frac{V_{j\rightarrow k}^{t}(\mathbf{x}_j)}{V_{j\rightarrow k}^{t}(\mathbf{\tilde{x}}_j)}\\
&= \text{log}\frac{p(\mathbf{x}_j|L^{s,p}(\mathbf{x}_j))\prod_{s\in{\partial j}\backslash k}{U_{s\rightarrow j}^{t-1}(\mathbf{x}_j)}}{p(\mathbf{\tilde{x}}_j|L^{s,p}(\mathbf{\tilde{x}}_j)\prod_{s\in{\partial j}\backslash k}{U_{s\rightarrow j}^{t-1}(\mathbf{\tilde{x}}_j)}}\\
&=L^{s,p}(\mathbf{x}_j)+\sum_{s\in{\partial j}\backslash k}LU_{k\rightarrow j}^{t-1}(\mathbf{x}_j)\text{,}
\end{split}
\end{equation}
\begin{equation}\label{LU}
\begin{split}
LU_{k\rightarrow j}^{t}&(\mathbf{x}_j) = \text{log}\frac{\sum_{(\mathbf{x}_p)\partial k \backslash j}\text{exp}[f_k(\mathbf{x}_j)]\prod_{p\in{\partial k}\backslash j}V_{p\rightarrow k}^{t}(\mathbf{x}_p)}{\sum_{(\mathbf{x}_p)\partial k \backslash j}\text{exp}[f_k(\mathbf{\tilde{x}}_j)]\prod_{p\in{\partial k}\backslash j}V_{p\rightarrow k}^{t}(\mathbf{x}_p)}\\
&=\text{log}\frac{\sum_{(\mathbf{x}_p)\partial k \backslash j} \text{exp}[f_k(\mathbf{x}_j)+\sum_{p\in{\partial k}\backslash j}LV_{p\rightarrow k}^{t}(\mathbf{x}_p)]}{\sum_{(\mathbf{x}_p)\partial k \backslash j}\text{exp}[f_k(\mathbf{\tilde{x}}_j)+\sum_{p\in{\partial k}\backslash j}LV_{p\rightarrow k}^{t}(\mathbf{x}_p)]}\text{.}
\end{split}
\end{equation}

In equation (\ref{LU}), $f_k(\mathbf{x}_j)=\frac{1}{N_0}\parallel y_k-h_{k,j}x_{k,j}\nonumber-\sum_{p\in\partial k \backslash j}{h_{k,p}x_{k,p}}\parallel ^2$. And in the final round, the output of SCMA detector is
\begin{equation} \label{scma_output}
LV_j(\mathbf{x}_j)=L^{s,p}_{j}(\mathbf{x}_j)+\sum_{s\in{\partial j}}LU_{k\rightarrow j}^{T}(\mathbf{x}_j)\text{.}
\end{equation}
Hence, the output can be decomposed into two parts: one is the prior information and another is the intrinsic information derived from the structure of SCMA factor graph. From another perspective, the LLR of $\mathbf{x}_j$ given prior information can be written as
\begin{equation}
\begin{split}
L^s_j&(\mathbf{x}_j)=\text{log}\frac{p(\mathbf{x}_j|\mathbf{y},\{L^{s,p}(\mathcal{X}^j)\}_{j=1,\cdots,J})}{p(\mathbf{\tilde{x}}_j|\mathbf{y},\{L^{s,p}(\mathcal{X}^j)\}_{j=1,\cdots,J})}\\
&=\text{log}\frac{p(\mathbf{y}|\mathbf{x}_j,\{L^{s,p}(\mathcal{X}^j)\}_{j=1,\cdots,J})}{p(\mathbf{y}|\mathbf{\tilde{x}}_j,\{L^{s,p}(\mathcal{X}^j)\}_{j=1,\cdots,J})}
+\text{log}\frac{p(\mathbf{x}_j|L^{s,p}(\mathbf{x}_j))}{p(\mathbf{\tilde{x}}_j|L^{s,p}(\mathbf{\tilde{x}}_j))}\\
&=L^{s,i}_{j}(\mathbf{x}_j)+L^{s,p}_{j}(\mathbf{x}_j)\text{.}
\end{split}
\end{equation}
It is apparent that $L^s_j(\mathbf{x}_j)$ also consists of two parts, the intrinsic parts and prior parts. when using MPA SCMA detection, we apply the term $\sum_{s\in{\partial j}}LU_{k\rightarrow j}^{T}(\mathbf{x}_j)$ to approximate $L^{s,i}_{j}(\mathbf{x}_j)$.
\subsection {LDPC decoding}
In this section, we briefly introduce the decoding process of LDPC for the sake of clear description of the combination of SCMA detection with LDPC decoder in the next section.

For LDPC is a sparse liner block code, it can be effectively expressed as factor graph which is a bipartite graph with check nodes and variable nodes. Take into consideration the complexity of implementation, LDPC decoder usually adopts BP algorithm using LLRs to reduce multiplication and avoid normalization. During each iteration, belief information is exchanged between variable nodes and check nodes based on check matrix. Let $L_{1,j}^{b,p}(b_i)$ be the LLR of the $i$-th bit of user $j$ input to the decoder, i.e., $L_{1,j}^{b,p}(b_i)=\text{log}\frac{p(b_i=0|\mathbf{y})}{p(b_i=1|\mathbf{y})}$, then the output of LDPC decoder is
\begin{equation}
L_{2,j}^{b}(b_i)=L_{1,j}^{b,p}(b_i)+\sum_{m\in\partial i}L_{j,m\rightarrow i}\text{,}
\end{equation}
where the notation $\partial i$ denotes the set of parity check functions variable $i$ belongs to and $L_{j,m\rightarrow i}$ is the belief information passed from check node $m$ to variable node $i$ in the final iteration. And the summation $\sum_{m\in\partial i}L_{j,m\rightarrow i}$ can be seen LDPC intrinsic information $L_{2,j}^{b,i}(b_i)$ based on LDPC structure. Hence, the output of LDPC decoder can also be decomposed to intrinsic information parts and prior information parts. Note that to distinguish the information generated by SCMA detector and LDPC decoder, we use subscript ``1'' and ``2'' respectively.

In the following section, we'll cope with some obstacle to implement the novel combination of SCMA detection and LDPC decoding and present the iterative scheme.
\begin{figure}[t]
\subfigure[Connection module for user $j$.]{
\label{fig3:periodCompare}
\centering
\includegraphics[width=3.2in]{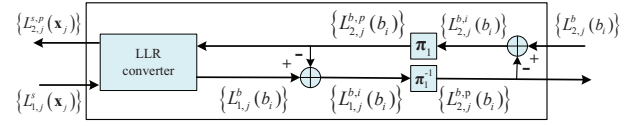}
}
\hspace{-5mm}
\subfigure[The information flows.]{
\label{a}
\centering
\includegraphics[width=3.2in]{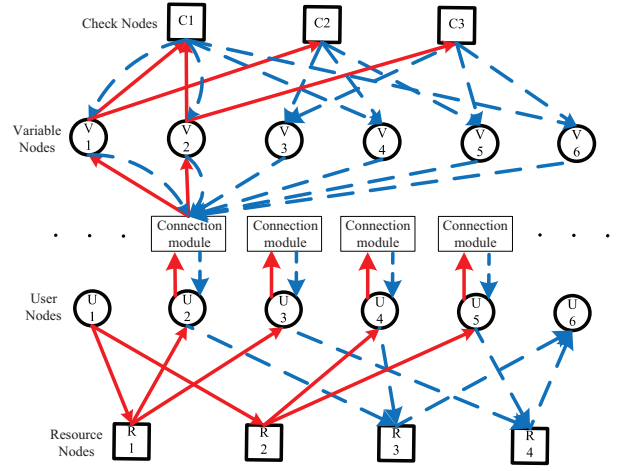}
}
\caption{The connection module and the information flows containing a specific symbol information from user node 1 to user node 6 in the combination scheme where 6 users and 4 resources exit and 1/2 rate LDPC code is adopted. The red arrow line denotes the flow of information given out from user 1 and the blue dashed arrow line denotes the flow of information feed back from LDPC decoder. Here, for clarity, we assume $\text{I}_{T}=\text{I}_{L}=1$, a symbol is formed by two bits and only the flows in LDPC factor graph of user 2 is drawn.}
\vspace{-6mm}
\label{information_flow}
\end{figure}

\subsection {Combination of LDPC decoding and SCMA detection}
From equation (\ref{scma_output}), the output of SCMA detector is at symbol level, but unfortunately the bit LLRs are required for input of LDPC decoder. In order to do a combination of LDPC decoding and SCMA detection, one problem we should deal with is the transformation between symbol LLRs and bit LLRs. The input to LDPC decoder should be intrinsic information parts of the multiuser detector output for a better performance like the decoding method for Turbo codes. Then another problem arises that the prior information part for a symbol comprise several bits' prior information and for a specific bit of this symbol, the other bits' prior information can be seen as intrinsic information, so we should decompose the output of SCMA detector into two parts, prior bit LLR and intrinsic bit LLR.

Here we assume that an SCMA symbol consists of $n=\text{log}_2(M)$ bits. Without loss of generality, the interleaved coded bits for a symbol can be considered independent. For a specific symbol $\mathbf{x}_j$, let $\mathcal{X}_j^{+}=\{i~|\text{if the i-th bit of $\mathbf{x}_j$ is} ``1"\}$ and $\mathcal{X}_j^{-}=\{i~|\text{if the i-th bit of $\mathbf{x}_j$ is} ``0"\}$. Then, the symbol LLR can be calculated based on bit LLRs as
\begin{equation}
L^{s,p}_{2,j}(\mathbf{x}_j)=\sum_{i\in \mathcal{X}_j^{-}}L_{2,j}^{b,p}(b_i)\text{.}
\end{equation}
Let $\mathcal{X}_i^{j,0}$ and $\mathcal{X}_i^{j,1}$ denote the set of the symbols of which the $i$-th bit corresponding to ``0" and ``1" for user $j$, respectively. The bit LLR can be written as
\begin{equation}
\begin{split}
&L_{1,j}^{b}(b_i)=\text{log}\frac{\sum_{\mathbf{x}_j\in\mathcal{X}_i^{j,0}}\text{exp}[LV_j(\mathbf{x}_j)]}{\sum_{\mathbf{x}_j\in\mathcal{X}_i^{j,1}}\text{exp}[LV_j(\mathbf{x}_j)]}\\
&=\text{log}\frac{\sum_{\mathbf{x}_j\in\mathcal{X}_i^{j,0}}\text{exp}[L^{s,p}_{2,j}(\mathbf{x}_j)+\sum_{s\in{\partial j}}LU_{k\rightarrow j}^{T}(\mathbf{x}_j)]}{\sum_{\mathbf{x}_j\in\mathcal{X}_i^{j,1}}\text{exp}[L^{s,p}_{2,j}(\mathbf{x}_j)+\sum_{s\in{\partial j}}LU_{k\rightarrow j}^{T}(\mathbf{x}_j)]}\\
&=L_{1,j}^{b,i}(b_i)+L_{2,j}^{b,p}(b_i)\text{.}
\end{split}
\end{equation}
The last equation follows the fact that $L^{s,p}_{2,j}(\mathbf{x}_j)$ is a combination of $\{{L_{2,j}^{b,p}(b_i)}\}_{i=1, \cdots, n}$ and we can extract a same product term $\text{exp}[L_{2,j}^{b,p}(b_i)]$ out of each sum term in the numerator.

Let $\text{I}_{T}$ and $\text{I}_{L}$ be the number of iterations of SCMA detector and LDPC decoder, respectively. The scheme of the novel iterative combination is described as follows. Besides outer iterations, this scheme consists of two inner stages: an SCMA detection stage with $\text{I}_{T}$ iterations, followed by LDPC decoding stage with $\text{I}_{L}$ iterations and the number of inner iterations of both stages can be much less than the number of iterations required by the SCMA receiver with or without traditional Turbo-like outer iterations. The two inner stages are connected by a connection module containing LLR converters, interleavers and deinterleavers. During each outer iteration, the symbol LLRs $\{L^{s,p}_{1,j}(\mathbf{x}_j)\}$ output by SCMA detector are converted into bit LLRs $\{L_{1,j}^{b}(b_i)\}$ by LLR converter.In order to obtain a better performance, the intrinsic bit LLRs $\{L_{1,j}^{b,i}(b_i)\}$ from SCMA detector should be extracted and then feed to the deinterleaver to get $\{L_{1,j}^{b,p}(b_i)\}$ which is fed into LDPC decoder as prior information. The process to feed information from LDPC decoder output to SCMA detector is similar. During each outer iteration, the detector and decoder only do a few number of iterations, less than the receiver without Turbo-like scheme need, although the total number of iterations may be the same. We call this characteristics partial inner iterations. For example, numerical results in the next section show that to obtain a satisfied BER performance 8 iterations are needed in traditional SCMA receiver, but 2 inner iterations are enough in our scheme with 4 outer iterations and the BER performance is even better. Hence this iterative combination can achieve a more satisfied BER performance gain with almost the same complexity as shown in the next section.
The connection module and the information flow of a special case where $\text{I}_{T}=\text{I}_{L}=1$ is shown as Fig.\ref{information_flow}. We call this special case mode 1. In particular, in mode 1 the detector does 1 demapping iteration and then the LLR goes to LDPC decoder through connection modules. The LDPC decoder only does 1 iteration and fed back the LLR to the detector.  Then we goes to the next iteration between detector and LDPC decoder. The best number of inner iterations under the constraint of total iteration number is unsolved and will be our future work. Intuitively, the more outer iterations the more intrinsic information exchanged, and hence the better performance. However, it will be shown in the next section that when take the extreme case where $\text{I}_{T}=\text{I}_{L}=1$ the BER performance is poor. One reason may lie in the fact that if $\text{I}_{T}=1$, it can be seen from Fig.\ref{information_flow} that the information interacts in SCMA detector is far from sufficient, i.e., there is no red arrow line connecting user node 1 with user node 6. Hence, in this scenario the information given out from SCMA detector is particularly poor so that the information cannot be corrected by LDPC decoder with only 1 iteration.

\subsection {Complexity analysis}
In this section, considering regular SCMA systems, we evaluate the complexity of traditional scheme for LDPC-coded SCMA receiver and our proposed scheme, i.e., the average number of operations we need for each symbol. Since the main complexity is introduced by multiplication, division, logarithm and exponent operations, we mainly consider these four kinds of operations for real number. From equation (\ref{LV}) and (\ref{LU}), it is clear that each inner iteration of SCMA detection needs $(2d_kKM^{d_k}+4{d_k}^2KM)/J$ multiplication operations, $d_kKM^{d_k}/J$ division operations $d_kKM^{d_k}/J$ exponent operations and $KMd_k/J$ logarithm operations. For regular LDPC codes with 1/2 code rate, $(11P-9)\text{log}_2(M)$ multiplications and $(P+1)\text{log}_2(M)$ divisions are needed during each iteration of decoding, \cite{fossorier1999reduced}, where P denotes the degree of variable nodes. For other kind of code rates, the order of complexity is the same \cite{mackay1999good}. $M\text{log}_2(M)$ exponent operations and $2\text{log}_2(M)$ logarithm operations are needed in connection module during each outer iterations and these operations are caused by the converting from symbol LLRs to bit LLRs.
It is now obvious that the main complexity is introduced by SCMA detection. Let $\text{I}_{O}$ be the number of outer iterations. From the deduction above and ignore the lower oder terms, it is easy to get that the total number of multiplication operations is approximately equal to $2I_OI_Td_kKM^{d_k}/J$, division and exponent operations are both approximately equal to $I_OI_Td_kKM^{d_k}/J$.

Here, with a slight abuse of notation, let the superscript t and p denote traditional scheme and proposed scheme, respectively. For the traditional scheme $I^t_O=1$, and for the proposed scheme $I^p_OI^p_T$ can be approximately equal to $I^t_T$. Hence, the complexity of our proposed scheme is almost the same as traditional receiver for LDPC-coded SCMA systems.
\section{Numerical results}
In this section, numerical simulations are conducted to evaluate the performance of our proposed scheme. This simulation is based on the LDPC codes with finite block size. The regular systematic LDPC code in China Mobile Multimedia Broadcasting (CMMB) systems is used in the simulations \cite{hu2013low}. The code length we adopted is 9216 bits and code rate is $R=1/2$. Bit error rates (BER) over AWGN channel and Rayleigh fading channel with $J=6,~K=4,~M=4$, and $N_j=2$ for all $j$ are presented. The user-specific codebooks are designed according to \cite{taherzadeh2014scma}. The channel coefficient $h$ is set to unit power for both AWGN channel and Rayleigh fading channel, i.e., $E(h^2)=1$. To keep numerically stable, when computing the term like $\text{log}(e^a+e^b)$, we adopt the equation as $\text{log}(e^a+e^b)=\text{max}(a,b)+\text{log}(1+e^{-|a-b|})$ \cite{hoshyar2008novel}.
\begin{figure}[t]
            \centering
            \includegraphics[width=3.2in]{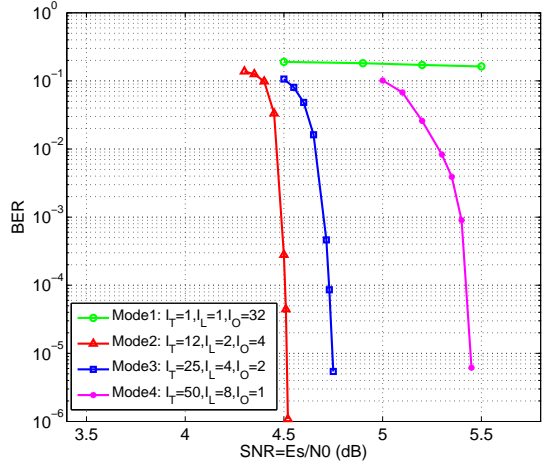}
            \hspace{2in}\parbox{1\linewidth}
{\caption{The BER performance over AWGN channels. }
\label{AWGN}}
            \end{figure}

\begin{figure}[t]
            \centering
            \includegraphics[width=3.2in]{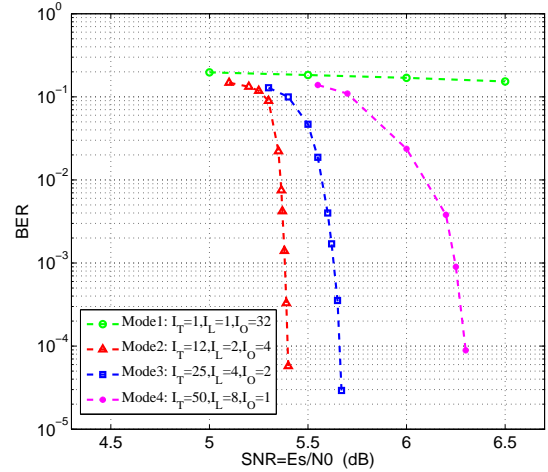}
            \hspace{2in}\parbox{1\linewidth}
{\caption{The BER performance over Rayleigh fading channels. }
\label{Rayleigh}}
            \end{figure}

Here, we compare four different modes, of which the total number of iterations is almost the same excepting the mode 1 ``$\text{I}_{T}=\text{I}_{L}=1, \text{I}_{O}=32$". The purpose to compare this mode with great difference in total iteration number is to show that the BER performance of this extreme case deteriorates seriously as shown in Fig. \ref{AWGN} and Fig. \ref{Rayleigh}. This poor BER performance of Mode 1 is consistent with the analysis in the last section. Fig. \ref{AWGN} shows that the required ${E_s/N_0}$ of the Mode 2 to achieve a BER of $10^{-4}$ is 4.5 dB on AWGN channels and 5.4 dB on Rayleigh fading channels, obtaining 0.3 dB gain over Mode 3 and 0.9 dB gain over Mode 4 on both kinds of channels. Note that Mode 4 is the traditional SCMA receiver scheme, i.e., there is no information fed back from decoder to detector. It can also be recognized that the BER \textit{waterfall} region of Mode 3 is more narrow than Mode 4 and that of Mode 2 is even more narrow than Mode 3. The phenomenon that a better BER performance and more narrow \textit{waterfall} region of Mode 2 and Mode 3 than Mode 4 owes to the ability of iterative combination that more \textit{diversity} gain can be achieved. Note that the complexity of the last three modes is almost the same if ignore the complexity caused by interleaver. If let Mode 2 and Mode 4 have the same BER performance, Mode 2 should reduce the number of iterations. Hence the proposed iterative scheme can be viewed as a method to reduce complexity from another perspective.

\section{Conclusions}
In this paper, we have proposed a novel Turbo-like combination scheme of SCMA detector and LDPC decoding with almost the same complexity compared to traditional SCMA receiver (non-iterative structure). Firstly, we have investigated the SCMA detection algorithm based on MPA from the aspect of solving marginal functions and shown this MPA-based detection is an approximation of MAP criterion since the factor graph of SCMA is usually not cycle-free. Then the MPA-based SCMA detection in Logarithmic form as well as the intrinsic information of the output have also been deduced in order to apply a novel soft Turbo-like combination with LDPC decoder. The combination scheme is described in detail and the information flows of a special case are presented. Numerical results show that the proposed scheme can achieve an excellent performance gain and more narrow waterfall region in terms of BER compared with traditional SCMA receiver for both AWGN channels and Rayleigh fading channels.

\bibliographystyle{IEEEtran}
\bibliography{SCMA_iterative}
\end{document}